\documentclass[showpacs,twocolumn,amsmath,amssymb,prb]{revtex4-1}
\usepackage{graphicx}
\usepackage{bm}
\usepackage{epsfig}
\usepackage{color}
\usepackage[]{natbib} 
\usepackage{url}
\def\rr{{{\bf r}}}
\def\rrp{{{\bf r}^\prime}}

\begin{document}
\title{Density-functional theory study of ionic inhomogeneity in metal clusters using SC-ISJM}

\author{M. Payami}
\affiliation{Theoretical and Computational Physics Group, NSTRI, AEOI, P.~O.~Box~14395-836, Tehran, Iran}
\email[Corresponding author: ]{mpayami@aeoi.org.ir}
\author{T. Mahmoodi}
\affiliation{Department of Physics, Mashhad Branch, Islamic Azad University, Iran}

\date{\today}
\begin{abstract}
In this work we have applied the recently formulated self-compressed inhomogeneous stabilized jellium model [M. Payami and T. Mahmoodi, Can. J. Phys. {\bf 89}, 967 (2011)] to describe the equilibrium electronic and geometric properties of atomic-closed-shell simple metal clusters of Al$_N$ ($N$=13, 19, 43, 55, 79, 87, 135, 141), Na$_N$, and Cs$_N$ ($N$=9, 15, 27, 51, 59, 65, 89, 113). To validate the results, we have also performed first-principles pseudo-potential calculations and used them as our reference.
In the model, we have considered two regions consisting of ``surface'' and ``inner'' ones, the border separating them is sharp. This generalization makes possible to decouple the relaxations of different parts of the system.
The results show that the present model correctly predicts the size reductions seen in most of the clusters. It also predicts increase in size of some clusters, as observed from first-principles results. The other property is the change of inter-layer distances, which in most cases were realized as contractions and in a few cases as expansions, are reproduced in good agreement with the atomic simulation results. For a better description of the properties, it is possible to improve the method of choosing the thicknesses or generalize the model to include more regions than just two.  
\end{abstract}
\maketitle
\section{\label{sec1}Introduction}
Atomic clusters have attracted a great deal of attention in the past three decades because of their unique properties different from those of single atoms and bulk materials.\cite{deHeer87,Jena87,Jena92,Haberland94,Ellis95,Jena96,Martin96,Sugano98,Jellinek99,Ekardt99,MeiwesBroer00,ReinhardSuraud04,Alonso05,Connerade08,philip2012evolution,stambula2014chemical,ditmire2016high,zhang2016creating,verkhovtsev2017irradiation} The physical properties of clusters depend on the atomic content and the geometry as well as the number of constituent atoms $N$. Nowadays, due to the significant progress in the experimental methods, it is possible to synthesize clusters with some desired properties. Among the various types of clusters, metal clusters (MC) are of much interest, firstly because of their biological and technological applications,\cite{LangBernhardt,Wangetal2013,Yinetal,YinBernstein,Yudanovetal,Collins,Shang,Yizhong,Shimojo,ChenChoiKamat,Bonacic-Koutecky} and secondly because the theoretical description of metallic systems are relatively simple. 

In fact, with the first-time production of free alkali clusters,\cite{Knight84} it was revealed that the quantized motion of the delocalized valence electrons in the average potential created by the ions is dominantly responsible for the shell effects observed in the abundance spectra. That is, the properties of alkali and other simple metal clusters are not much sensitive to the detailed structure of the ions.\cite{deHeer93} This fact led to the use of simple model, called jellium model (JM), in which the discrete ionic cores are replaced with a uniform positive background.\cite{Ekardt84a,Ekardt84b,Brack93} 

Historically, development of the JM aimed to describe physical properties of metals within a small computational effort. It was extensively shown that this model was successful in describing some properties of simple s-p bonded metals.\cite{AshcroftLangreth67,LangKohn70,LangKohn71,PerdewMonnier76,PerdewMonnier80,Perdew82,Kiejna-Woj96} However, rapid increasing of computational power along with the development of accurate quantum-chemical {\it ab initio} methods did not discourage researchers from finding simple models for accurate description of the materials' properties. In contrast to the use of quantum-chemical {\it ab initio} methods which are appropriate for the accurate description of only small clusters (containing at most a few hundred atoms), using the JM enables one to go far beyond this limitation and describe the properties of very large simple MCs qualitatively and semi-quantitatively within an acceptable time period. In the first-time JM description of metal clusters, the positive ionic background was replaced by a jellium sphere of uniform charge in the electrostatic field of which the interacting valence electrons were moving. In subsequent works, spheroidal and ellipsoidal shapes for the jellium were considered which were useful in describing the odd-even effects seen experimentally. In one of the works, to describe the observed hollow clusters, the JM was applied using spherical shells of uniform charge.\cite{Polozkov09} 

Although the simple JM was successful in describing some effects observed in metal clusters, it was unable to predict accurate values for the energetics. It was also shown that simple JM, because of its mechanical instability, would lead to some inaccurate results for systems of very high or very low electron densities.\cite{LangKohn70,AshcroftLangreth67} 

Introducing the stabilized jellium model (SJM) was the first successful attempt\cite{Perdew90} to overcome the above-mentioned drawbacks of the JM and get some accurate results for the bulk properties. In the SJM, the observed bulk density parameter, $r_s^B$, was used as input parameter for the background jellium density. 

In applying the SJM for finite systems, because of the ionic relaxation near the surface, the $r_s^B$ would not bring the system to mechanical stability. In that case, the appropriate parameter, $r_s^\dagger$, which is close to $r_s^B$ for sufficiently large system, minimizes the total energy and brings the system to mechanical stability. The method was called self-compressed SJM (SC-SJM),\cite{Perdew93,Payami01,Payami04,Payami06,Mahmoodi09} which could only predict an overall expansion or contraction for the whole system, and was not capable to explain the ionic relaxation in the surface region. 

To explain the properties near the surface of a system, one should take a special care for the surface region, and to this end it is natural to assume that the whole system may be divided into two "surface" and "bulk" regions with two different jellium densities (the boundaries between regions are taken to be sharp), $\bar{n}_1$ and $\bar{n}_2$, as independent variables for the inner bulk and surface regions, respectively. This generalized model which was called the self-compressed inhomogeneous stabilized jellium model (SC-ISJM) has recently been formulated by the authors and successfully applied to describe the surface effects in simple metal thin films.\cite{PayamiMahmoodi11} 

In this work, we have applied the SC-ISJM to determine the surface- and bulk-region relaxations for closed-atomic-shell $N$-atom clusters of Al$_N$ ($N$=13, 19, 43, 55, 79, 87, 135, 141) as well as Na$_N$ and Cs$_N$ ($N$=9, 15, 27, 51, 59, 65, 89, 113).   
Our results show that for most of them, the surface region contracts to higher electron and ionic densities, in good agreement with the first-principles method results. The inner regions for some of them expand to lower electron and ionic densities, and for others it contracts. That is, some clusters in their equilibrium states tend to have a relatively hollow shapes than being uniformly distributed over the space. 
Also, our calculations for Na$_{65}, $Cs$_{51}$ and Cs$_{65}$ clusters show that, in contrast to others, they undergo surface expansions. The results for Na$_{65}$ and Cs$_{65}$ are also in good agreement with those obtained from first-principles pseudo-potential method.\cite{QE} Comparison of the two sets of results reveals that the present model, in which the ionic constraints are somewhat released, give lower energies for the ground states and is more realistic than previously established ones. 

The organization of this paper is as follows.       
In section \ref{sec2}, the SC-ISJM is casted in the form suitable for clusters. Section \ref{sec3} is dedicated to the calculation details, and the results are discussed in section \ref{sec4}. Finally, in section \ref{sec5}, we conclude this work.         
%
\section{Formulation of SC-ISJM for metal clusters}\label{sec2}
The total energy of a system in the SC-ISJM is given by \cite{PayamiMahmoodi11}
\begin{widetext}
\begin{eqnarray}\label{eq1}
\nonumber E_{\rm ISJM}\left[n;[\{n_+^{(\alpha)}\}]\right]&=&E_{\rm JM}\left[n;[\{n_+^{(\alpha)}\}]\right]+\sum_{\alpha}\langle\delta v\rangle_{\rm WS}(\bar{n}_\alpha)\int d\rr~\Theta^{(\alpha)}(\rr)~[n(\rr)-n_+^{(\alpha)}(\rr)] \\
                &+&\sum_\alpha [\bar{w}_R(\bar{n}_\alpha)+\varepsilon_{\rm M}(\bar{n}_\alpha)]\int d\rr~n_+^{(\alpha)}(\rr),
\end{eqnarray}
\end{widetext}
where,
\begin{widetext}
\begin{equation}\label{eq2}
E_{\rm JM}\left[n;[\{n_+^{(\alpha)}\}]\right]=T_s[n]+E_{xc}[n]    
       +\frac{1}{2}\int~d\rr~\phi([n,n_+];\rr)~[n(\rr)-n_+(\rr)]
\end{equation}
\end{widetext}
is the simple-JM energy, in which $T_s$ and $E_{xc}$ being the non-interacting kinetic and exchange-correlation energies, respectively; and the last term in the right hand side, the electrostatic energy, is given by
\begin{equation}\label{eq3}
\phi([n,n_+];\rr)=\int d\rrp~\frac{[n(\rrp)-n_+(\rrp)]}{|\rr-\rrp|}
\end{equation}
with the positive charge distribution 
\begin{equation}\label{eq4}
n_+(\rr)=\sum_\alpha n_+^{(\alpha)}(\rr);\,\,\,n_+^{(\alpha)}(\rr)\equiv \bar{n}_\alpha\Theta^{(\alpha)}(\rr),
\end{equation}
in which $\Theta^{(\alpha)}(\rr)$ takes the value of unity inside the region $\alpha$, and zero outside. $\bar{n}_\alpha$ is the constant jellium density of region $\alpha$. All equations throughout the paper are expressed in Hartree atomic units. 

In equation ~(\ref{eq1}), the averaged difference potential is given by
\begin{equation}\label{eq5} 
\langle\delta v\rangle_{\rm WS}(\bar{n}_\alpha)=\frac{1}{\Omega^{(\alpha)}}\int_{\Omega^{(\alpha)}} d\rr\;\delta v(\rr);\;\;\;\;\Omega^{(\alpha)}=\frac{z}{\bar{n}_\alpha},
\end{equation}
where $\delta v$ is the difference potential between the pseudo-potential\cite{Ashcroft66} of ions and the electrostatic potential of the jellium background.   
$\bar{w}_R$ and $\varepsilon_{\rm M}$ are the average of the repulsive part of the pseudo-potential and the Madelung energy, respectively.

To determine the equilibrium state of the system with specified regions, the following minimization rule is applied: 
\begin{widetext}
\begin{equation}\label{eq6}
E_0\equiv E\left[n_0;[\{n_+^{(\alpha)\dagger}\}]\right]= \min_{\{n_+^{(\alpha)}\}} E\left[n_0;[\{n_+^{(\alpha)}\}]\right] =\min_{\{n_+^{(\alpha)}\}} \left\{ \min_n E\left[n;[\{n_+^{(\alpha)}\}]\right] \right\},
\end{equation}
\end{widetext}
in which the inner minimization after the last equality sign gives the ground-state energy of the system (with $n_0$ as the ground-state electron density) constrained to the external parameters $\{n_+^{(\alpha)}\}$, and the outer minimization determines the equilibrium distribution of positive charges, $\{n_+^{(\alpha)\dagger}\}$, in the system. 
 
\begin{figure}[h]
\begin{center}
\includegraphics[width=8.6cm]{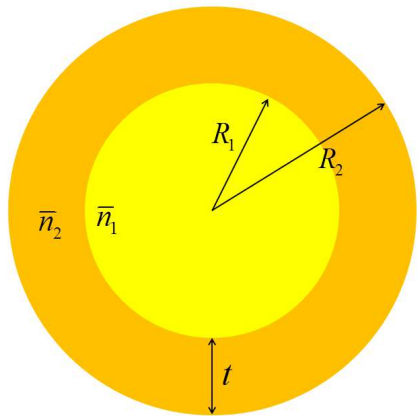}
\caption{\label{fig1} Spherical jellium with surface and inner regions. Inner region is specified by a sphere of uniform density $\bar{n}_1$ and radius $R_1$, while the surface region is confined between two spheres of radii $R_1$ and $R_2$ having uniform density of $\bar{n}_2$ and thickness $R_2-R_1=t$.}
\end{center}
\end{figure}

\begin{figure}[h]
\begin{center}
\includegraphics[width=8.6cm]{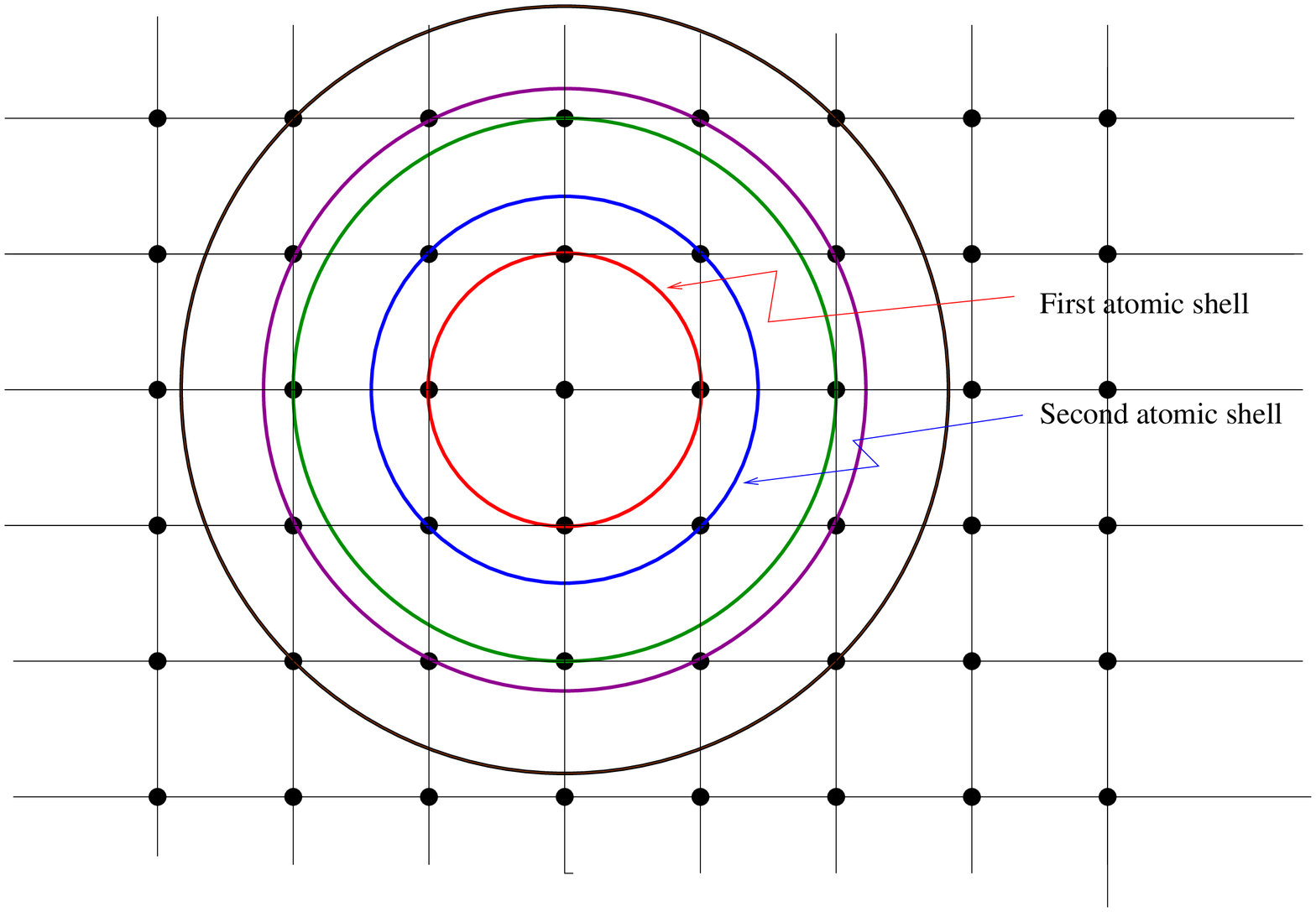}
\caption{\label{fig2} Schematic view of a two-dimensional unrelaxed 25-atom cluster composed of a central atom and 5 closed atomic shells, cut out of a two-dimensional cubic lattice.}
\end{center}
\end{figure}

\begin{figure}
\begin{center}
\includegraphics[width=8.6cm]{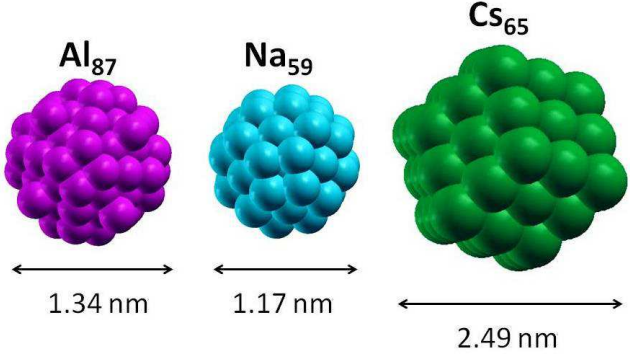}
\caption{\label{fig3} Closed-atomic-shell clusters of Al$_{87}$, Na$_{59}$, and Cs$_{65}$ which are resulted from adding successive closed atomic-shells.}
\end{center}
\end{figure}

For metal clusters, $\alpha$ takes the two values of 1 and 2 for the inner and surface regions, respectively. Assuming spherical geometry (which is appropriate for closed-atomic-shell clusters), the inner region is a sphere with radius $R_1$ and uniform density $\bar{n}_1$, while the surface region is a shell of thickness $t$ which is confined between two concentric spheres of radii $R_1$ and $R_2$ and having uniform density $\bar{n}_2$. The situation is schematically shown in Figure~\ref{fig1}.
We may consider that the isolated cluster is cut out from an infinite bulk system. Before taking the cluster out of the bulk, the two above-mentioned regions have the same uniform bulk densities, $\bar{n}_1=\bar{n}_2=\bar{n}^B$, and the radius of the outer sphere is given by $R_2^B=(zN)^{1/3}r_s^B$ with $N$ being the number of constituent atoms of the cluster, $z$ being the valence of the atom, and $r_s^B=(3/4\pi\bar{n}^B)^{1/3}$. The surface region is specified by the parameter $t=(R_2^B-R_1^B)$ which could in principle take any continuous real number that should not be less than some critical value which corresponds to states with large repulsive forces between the ions. The starting structure of a closed-atomic-shell cluster is constructed from an infinite lattice by choosing a central atom and adding spherical shells of atoms successively. The first, second, etc., shells consist of first, second, etc., nearest-neighbor atoms, respectively, as shown in figure~\ref{fig2}.
The result for aluminum is Al$_N$ clusters with $N$=13, 19, 43, 55, 79, 87, 135, 141 atoms, whereas for sodium and cesium are Na$_N$ and Cs$_N$ clusters with $N$=9, 15, 27, 51, 59, 65, 89, 113 atoms. In figure~\ref{fig3}, we have shown the typical closed-atomic-shell clusters Al$_{87}$, Na$_{59}$, and Cs$_{65}$ resulted from the above procedure. 

In applying the SC-ISJM for clusters, the minimization rule, Eq.~(\ref{eq6}), now reduces to 
\begin{eqnarray}\label{eq7}
E_0&\equiv& E\left[n_0;[t^\dagger,\bar n_1^\dagger,\bar n_2^\dagger]\right] \nonumber \\ 
    &=&\min_t\left\{\min_{\bar n_1,\bar n_2}\left\{\min_n E\left[n;[t,\bar n_1,\bar n_2]\right] \right\}\right\},
\end{eqnarray}
in which $t^\dagger$, $\bar n_1^\dagger$, $\bar n_2^\dagger$ are the equilibrium surface thickness, inner- and surface-region jellium densities, respectively. The values of $r_s^B$, used here for Al, Na, and Cs are 2.07, 3.99, and 5.67 Bohrs.
\section{Calculation methods}\label{sec3}
The equilibrium states of closed-atomic-shell clusters are calculated using the SC-ISJM as well as the first-principles pseudo-potential method, which are explined as follows.
\subsection{SC-ISJM}
To find the equilibrium state of the cluster through Eq.~(\ref{eq7}), that is, minimization over a three-parameter space, we first take a trial surface thickness, $t$ (through which the two regions are specified), and calculate the ground-state energy for the given external parameters $t$, $\bar n_1^B$ and $\bar n_2^B$, using the density functional theory\cite{HohenbergKohn} (DFT) and the self-consistent solution of the Kohn-Sham (KS) equations\cite{KohnSham} within  the local-density approximation\cite{KohnSham} (LDA). The calculations then are repeated for different points in the \{$\bar n_1$, $\bar n_2$\} parameter space (keeping the number of atoms in each region as fixed) until a local minimum is achieved.\cite{Powell64} The global minimum-energy state is then selected from the calculated local-minimum states for different $t$ values, and thereby the equilibrium value, $t^\dagger$, is singled out. For $t$ we arbitrarily use the values obtained from $t=\nu d_{100}$ in which $\nu$ takes positive integer values and $d_{100}$ is the distance between two adjacent (100) planes, being 3.82, 4.05, and 5.72 for Al, Na, and Cs, respectively. 

The constraint, that keeps constant the number of atoms in each region during the variations of jellium densities, is given by

\begin{equation}\label{eq9}
\bar{n}^B V_\alpha^B=\bar{n}_\alpha V_\alpha=\bar{n}_\alpha^\dagger 
  V_\alpha^\dagger;\,\,\,\,\alpha=1,2
\end{equation} 
 in which $\bar{n}_\alpha^\dagger$ and $V_\alpha^\dagger$ are the density and volume of region $\alpha$ in the equilibrium state, respectively; while $\bar{n}_\alpha$ and $V_\alpha$ are the corresponding quantities in an arbitrary non-equilibrium state. $V_\alpha^B$ corresponds to the volume of region $\alpha$ in its bulk un-relaxed state (before cutting out the cluster from the infinite bulk material). The constraints are easily expressed in terms of the densities of the two regions and the radii before and after relaxation as

\begin{equation}\label{eq10}
\left(R_2^B\right)^3=\frac{3zN}{4\pi\bar{n}^B},
\end{equation}

\begin{equation}\label{eq11}
R_1^B=R_2^B-t,
\end{equation}

\begin{equation}\label{eq12}
\bar{n}^B \left(R_1^B\right)^3=\bar{n}_1 \left(R_1\right)^3,
\end{equation}

\begin{equation}\label{eq13}
\bar{n}^B \left[\left(R_2^B\right)^3-\left(R_1^B\right)^3\right]=\bar{n}_2 \left[\left(R_2\right)^3-\left(R_1\right)^3\right].
\end{equation}

Given the number of constituent atoms, $N$, and valence $z$, the outer bulk radius of the cluster, $R_2^B$, is determined from equation (\ref{eq10}). Choosing a proper value for $t$, the value of the bulk inner radius, $R_1^B$, is determined from equation (\ref{eq11}). The set of equations~(\ref{eq10})-(\ref{eq13}) completely specify the set of external parameters \{$t,\bar{n}_1,\bar{n}_2$\}, from which the electrostatic potential energy of an electron for an arbitrary state is given by:

\begin{widetext}
\begin{equation}\label{eq14}
V(r)=\left\{\begin{array}{lll}
-2\pi\left[\bar{n}_1(R_1^2-\frac{r^2}{3})+\bar{n}_2(R_2^2-R_1^2)\right] &,& r\le R_1 \\
-\left\{\frac{Nz}{R_2}+\frac{4\pi}{3}\left[R_1^3(\bar{n}_1-\bar{n}_2)\left(\frac{1}{r}-\frac{1}{R_2}\right) -\frac{\bar{n}_2}{2}\left(r^2-R_2^2 \right) \right]\right\} & ,&  R_1 \le r \le R_2 \\
-\frac{Nz}{r} & , &  r> R_2.  
\end{array}\right .
\end{equation}
\end{widetext}

In the SC-ISJM calculations, we assume a full rotational symmetry for the clusters, and therefore, the physical quantities will depend on the single radial variable $r$. Under this constraint, by taking the KS orbitals as a product of a radial function and spherical harmonics
\begin{equation}\label{eq15}
\Psi_i(\rr)=\psi(r)Y_{l,m}(\theta,\phi),
\end{equation} 
the KS equations reduce to one-dimensional eigenvalue equations:
\begin{equation}\label{eq16}
\hat{H}^{(l)}\psi_n^{(l)}(r)=\varepsilon_n^{(l)}\psi_n^{(l)}(r), \;\;\;\;\;n=1,2,\ldots,
\end{equation}
where,
\begin{equation}\label{eq17}
\hat{H}^{(l)}\equiv -\frac{1}{2}\frac{d^2}{dr^2}-\frac{1}{r}\frac{d}{dr}+V_{eff}([n,n_+];r)+\frac{l(l+1)}{2r^2}
\end{equation}
in which
\begin{eqnarray}\label{eq18}
V_{eff}([n,n_+];r)&=&\phi([n,n_+];r)+V_{xc}([n];r) \nonumber \\ 
 &+&\sum_{\alpha=1}^2 \langle\delta v \rangle_{\rm WS}(\bar{n}^\alpha)\Theta^{(\alpha)}(r).
\end{eqnarray}

In this study, we have considered the closed-atomic-shell clusters. This does not imply that the resultant cluster is an electronic closed-shell one. On the other hand, if at the starting point ($\bar n_1=\bar n_2=\bar n^B$) we choose the number of constituent atoms in such a way that results in an electronic closed-shell state, then as soon as $\bar{n}_\alpha\ne \bar{n}^B$, the hierarchy of electronic levels differs from that in the starting configuration, and therefore, the relaxed cluster may become an electronic open-shell one. Moreover, in some cases where $\varepsilon_{\rm HOMO}\approx \varepsilon_{\rm LUMO}$, achieving the self-consistent point in the iterative solution of the KS equations becomes problematic. The workaround we adopted was to add a tiny ``smearing'' factor, $\sigma$, for the levels. That is, we consider the occupation of all states by the Fermi-Dirac weight. The resulting spherical averaged electron density then reads as
\begin{equation}\label{eq19}
n(r)=2\times\frac{1}{4\pi}\sum_{l,n}(2l+1)f_{l,n}\left[\psi_n^{(l)}(r)\right]^2
\end{equation}
in which $f_{l,n}$ is the Fermi-Dirac function,
\begin{equation}\label{eq20}
f_{l,n}\equiv f(\varepsilon_n^{(l)},\mu)=\frac{1}{\exp\{(\varepsilon_n^{(l)}-\mu)/\sigma\}+1}.
\end{equation}
The chemical potential, $\mu$, which is close to the Fermi energy $\varepsilon_{\rm F}$ for a tiny $\sigma$ value, is determined from the solution of the equation
\begin{equation}\label{eq21}
Nz=2\sum_{l,n}(2l+1)f_{l,n}.
\end{equation} 
This procedure is equivalent to treating the electrons in a heat bath of temperature $\sigma/k_{\rm B}$ ($k_{\rm B}$ is Boltzmann constant), and then the variational quantity would be the free energy given by
\begin{widetext}
\begin{equation}\label{eq22}
F=E_{\rm ISJM}[n,\{n_+^{(\alpha)}\}]+2\sigma\sum_{l,n}(2l+1)\left[f_{l,n}\log{f_{l,n}} + (1-f_{l,n}) \log{(1-f_{l,n})}    \right]
\end{equation}  
\end{widetext}
in which the second term in the right hand side is the contribution of the entropy of the electronic energy levels. The Fermi-Dirac weight should also be included properly in the ``band energy'' (sum of KS eigenvalues) which is used in the calculation of the total energy.

To solve the KS equations for a cluster, we expand the KS orbitals in equation (\ref{eq16}), in terms of basis functions, $W_s^{(l)}(r)$, which are the normalized radial eigenfunctions of an infinite spherical potential well,
\begin{equation}\label{eq23}
V(r)=\left\{ \begin{array}{lll}
0 &;& r<R \\
\infty &;& r\ge R,
\end{array}
\right.
\end{equation}
and are given by spherical Bessel's functions as
\begin{equation}\label{eq24}
W_s^{(l)}(r)=\left\{R^3\left[j_{l+1}(\beta_{l,s})\right]^2/2 \right\}^{-1/2} j_l(\frac{\beta_{l,s}}{R}r),
\end{equation}
where, $\beta_{l,s}$ is the $s$th root of the spherical Bessel's function of order $l$, that satisfies the equation
\begin{equation}\label{eq25}   
j_l(\beta_{l,s})=0, \;\;\;\;l=0,1,2,\ldots ; s=1,2,3,\ldots,
\end{equation}
and the radius of the well, $R$, is so chosen that $R\approx R_2^B + 2\lambda_{\rm F}^B$ is satisfied ($\lambda_{\rm F}^B$ is the Fermi wavelength of the valence electrons). The basis functions satisfy the orthonormality condition
\begin{equation}\label{eq26}
\int_{r=0}^{R} dr\;r^2 W_s^{(l)}(r) W_{s^\prime}^{(l)}(r) = \delta_{s,s^\prime},
\end{equation}
and the eigenvalues of the infinite well are given by
\begin{equation}\label{eq27}
E_{l,s}=\frac{\beta_{l,s}^2}{2R^2}.
\end{equation}

Inserting the expansion
\begin{equation}\label{eq28}
\psi^{(l)}(r)=\sum_s a_s^{(l)} W_s^{(l)}(r)
\end{equation}
into equation (\ref{eq16}), we get the matrix elements as
\begin{eqnarray}\label{eq29}
\left[\hat{H}^{(l)} \right]_{s^\prime,s}&\equiv& \int_{r=0}^R dr\;r^2 W_{s^\prime}^{(l)}(r) \hat{H}^{(l)} W_{s}^{(l)}(r)\nonumber \\ &=&\frac{\beta_{l,s}^2}{2R^2}\delta_{s^\prime,s} +  \left[V_{eff} \right]_{s^\prime,s}
\end{eqnarray}
from which we solve the secular equation and obtain the KS eigenvalues and eigenfunctions.
\subsection{First-principles method}
For the First-principles calculations, we have employed super-cell method with Martyna-Tuckerman correction\cite{MT1999} for isolated systems as implemented in the Quantum-ESPRESSO code package.\cite{QE} The size of the cubic super-cell for each cluster is taken to be enough large suitable for an isolated cluster, and all calculations are done at the $\Gamma$ point. For the Al, Na, and Cs atoms, we have used the pseudo-potentials ``Al.pz-vbc.UPF'', ``Na.pz-n-vbc.UPF'', and ``Cs.pz-nc.UPF'', respectively.\cite{QE} The calculations are based on the above-mentioned norm-conserving pseudo-potentials with a cutoff of 40 Ry for the plane-wave basis set and using smearing with 0.02 Ry of broadening. The geometries are fully relaxed until the total force on the cluster is less than 10$^{-5}$ Ry/au. Because of the symmetry inherent in the atomic positions of the closed-atomic-shell clusters, the relaxations take place in the radial direction of the atomic shells, and therefore, the changes in the inter-shell spacings can be compared with the SC-ISJM results.        
%
%
\section{\label{sec4}Results and discussion}
In this section, we present the calculation results of first-principles method and of the SC-ISJM method, and then discuss how the SC-ISJM results are consistent with those of {\it ab initio} method.
\subsection{First-principles results}
In order to determine the finite-size effects on the clusters, we have first calculated the optimized lattice constants for bulk Al, Na and Cs crystals with fcc, bcc, and bcc structures, and obtained the values of 7.48, 7.79, and 11.75~a.u. for the equilibrium lattice constants, respectively. In the bulk calculations, 
the plane-wave basis set cutoffs for the wave functions and the density were chosen as 40 and 160 Ry, respectively. Also, for Brillouin-zone integrations, we used a mesh of size $20\times 20\times 20$ in the k space.
We have then constructed our un-relaxed clusters out of these bulk-optimized structures as closed-atomic-shell ones containing successive shells of first-nearest-neighbors, second-nearest-neighbors, and so on, from the central atom. These structures are highly symmetric ones which may correspond not to a bound state of global-minimum energy, but this approximation makes them suitable for comparison with spherical SC-ISJM results. 

\begin{figure*}[t]
\begin{center}
\includegraphics[width=0.8\textwidth]{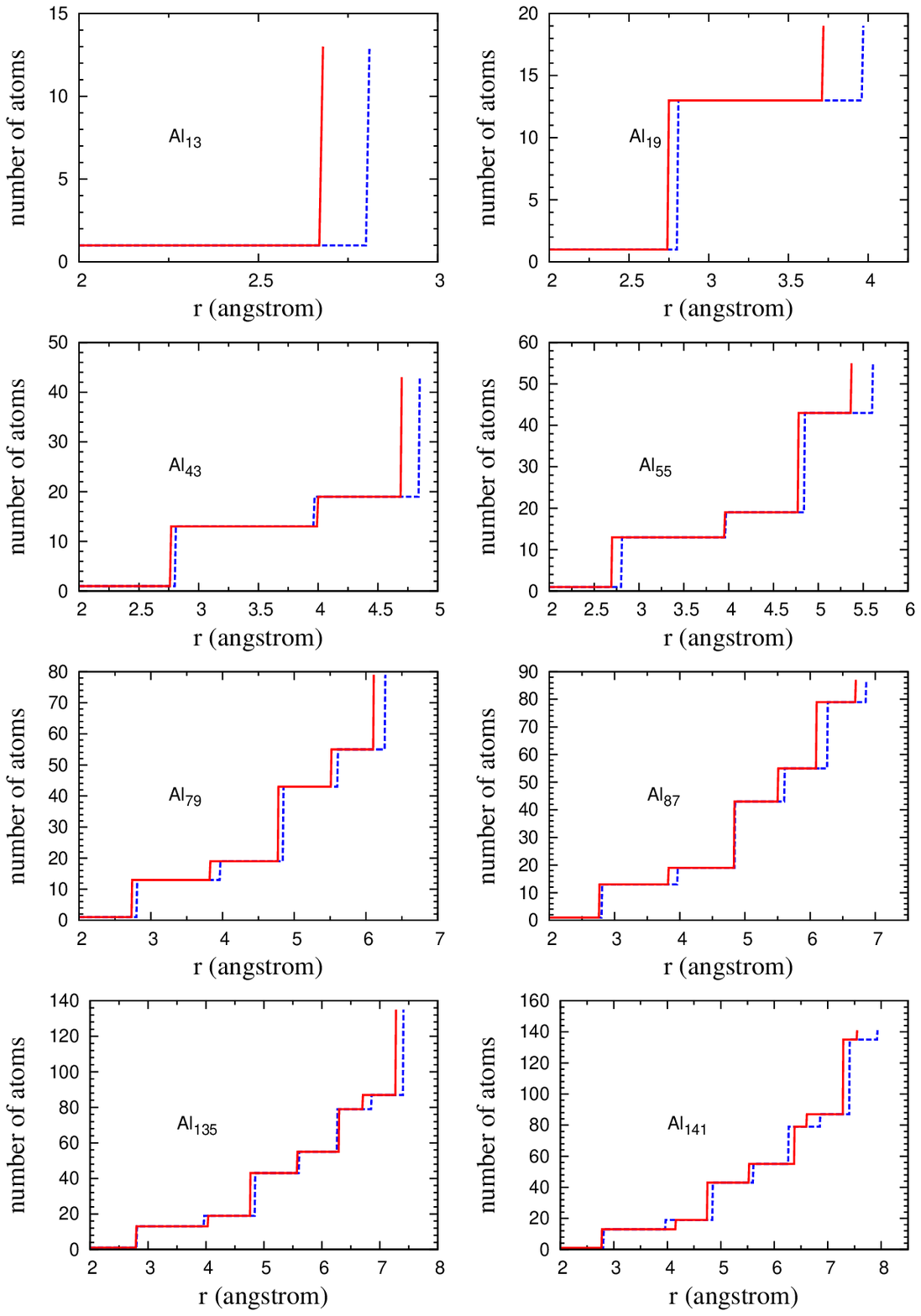}
\caption{\label{fig4} Number of atoms enclosed in shells of Al$_N$ clusters as functions of shell radii. Dashed blue lines correspond to un-relaxed clusters while the solid red lines represent the relaxed clusters.}
\end{center}
\end{figure*}

\begin{figure*}[t]
\begin{center}
\includegraphics[width=0.8\textwidth]{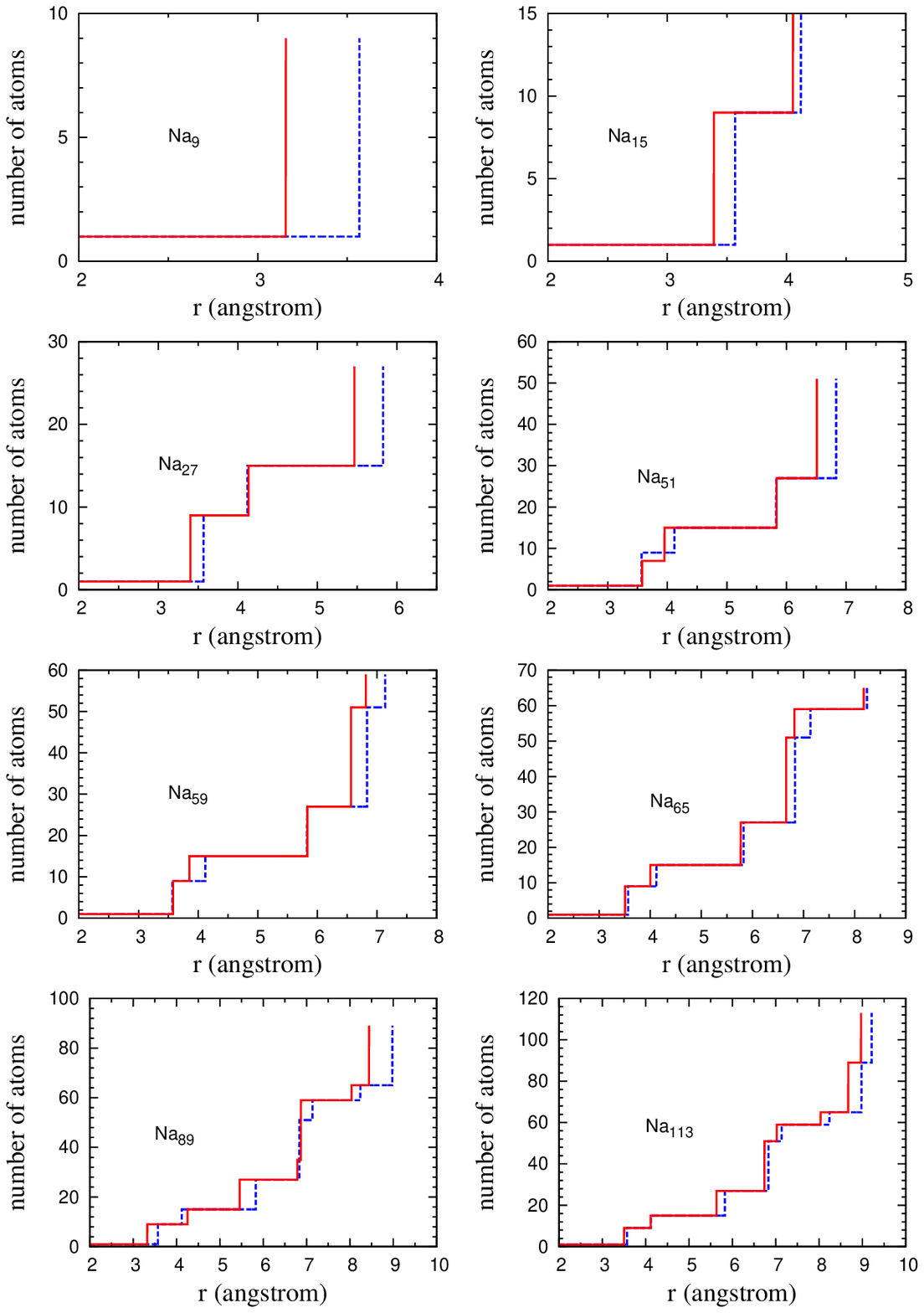}
\caption{\label{fig5} The same as in Fig.~\ref{fig4}, but for Na$_N$ clusters.}
\end{center}
\end{figure*}

\begin{figure*}[t]
\begin{center}
\includegraphics[width=0.8\linewidth]{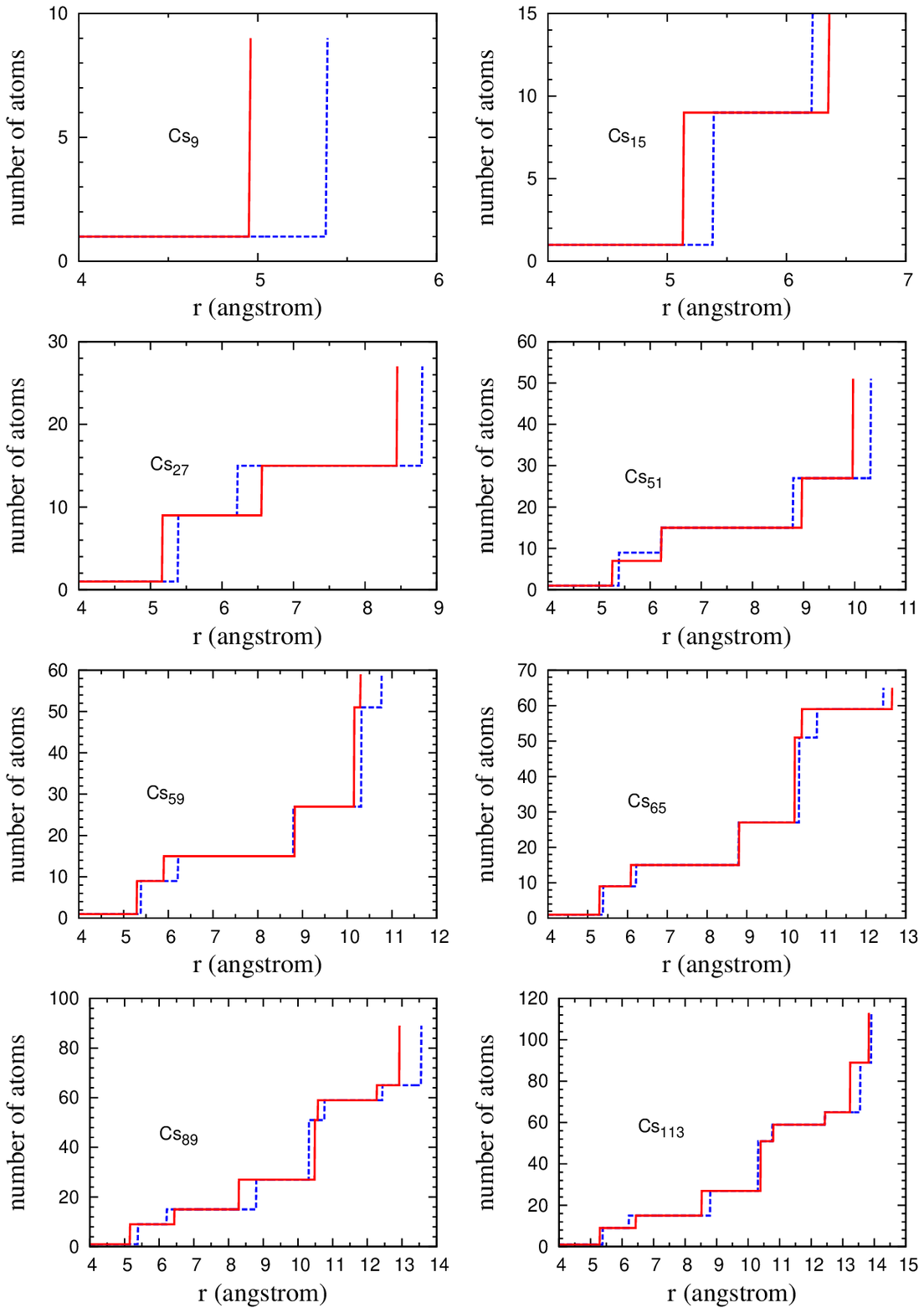}
\caption{\label{fig6} The same as in Fig.~\ref{fig4}, but for Cs$_N$ clusters.}
\end{center}
\end{figure*}

The radial positions and numbers of atoms enclosed by shells for the un-relaxed clusters are shown by dashed blue lines; and the optimized atomic configurations which preserve the $n^{th}$-nearest neighbor structures but with some new radial distances, are represented by solid red lines in Figs.~\ref{fig4}-\ref{fig6}. Looking at the radial distance of the outermost shell (which we call the first shell) for relaxed systems, we see that all clusters, but Cs$_{15}$ and Cs$_{65}$, undergo size reductions. For the mentioned two exceptions the sizes are increased. 

The results show that the second shells of all clusters; but Al$_{43}$, Na$_{27}$, Na$_{51}$, Cs$_{27}$, and Cs$_{51}$; were also contracted. As to the third shell, all were contracted except for Al$_{135}$, Na$_{59}$, Cs$_{51}$, Cs$_{59}$, and Cs$_{113}$ which were expanded or remained more or less unchanged. The behaviors of the innermost shells can also be inspected from the figures, but since we are mostly interested in the surface behaviors, we ignore those details and just mention their overall effects. 

The relative distances between the first and second layers were reduced for all except Al$_{87}$, Al$_{135}$, Na$_{15}$, Na$_{65}$, Na$_{113}$, Cs$_{15}$, Cs$_{65}$, Cs$_{113}$, and for these it is increased or remained unchanged even though the mean radii might be changed.              
The relative distance between the second and third layers show reductions for all except Al$_{43}$, Al$_{141}$, Na$_{27}$, Na$_{51}$, Na$_{89}$, Cs$_{27}$, Cs$_{51}$, Cs$_{89}$. In these exceptions it is increased or remained more or less unchanged.
Finally, the relative distances between the first and third layers were reduced for all except Na$_{65}$ and Cs$_{65}$. For these two exceptions, they were increased.  
\subsection{SC-ISJM results}
Using the SC-ISJM we have determined, through Eq.~(\ref{eq7}), the equilibrium surface region thickness, $t^\dagger=\nu^\dagger d_{100}$, and the corresponding equilibrium jellium densities, $\bar n_1^\dagger$, $\bar n_2^\dagger$, for Al$_N$ ($N$=13, 19, 43, 55, 79, 87, 135, 141), Na$_N$ and Cs$_N$ ($N$=9, 15, 27, 51, 59, 65, 89, 113) clusters. The results are summarized in Table~\ref{tab1}. 

\begin{table*}
\caption{\label{tab1}Ground state energies of Al$_N$, Na$_N$, and Cs$_N$ clusters resulted from SC-ISJM. Here, $\Delta r_s^\dagger = r_s^\dagger - r_s^B$. For comparison, the energies of SC-SJM and SJM are also listed. As is seen, in all cases the inequality $E_{\rm SC-ISJM} < E_{\rm SC-SJM} < E_{\rm SJM}$ holds.}
\begin{ruledtabular}\tiny
\begin{tabular}{ccccccccc}
    &    & \multicolumn{4}{c}{SC-ISJM}& \multicolumn{2}{c}{SC-SJM} & SJM \\ \cline{3-6} \cline{7-8} \cline{9-9}   
atom & $N$ & $\nu^\dagger$  &  $\Delta r_s1^\dagger$  & $\Delta r_s2^\dagger$  & $-E_{\rm SC-ISJM}$  &$\Delta r_s^\dagger$ & -$E_{\rm SC-SJM}$  &-$E_{\rm SJM}$ \\ \hline
Al &  13   &  1   & +0.205 & -0.209 & 27.1684   & -0.143& 27.1106 & 26.9684 \\
   &  19   &  1   & +0.185 & -0.101 & 39.7020   & -0.105& 39.5816 & 39.4659 \\
   &  43   &  1   & +0.206 & -0.169 & 90.0310   & -0.079& 89.6954 & 89.5462 \\
   &  55   &  1   & +0.206 & -0.068 & 114.7170  & -0.074& 114.6332&114.4648\\
   &  79   &  1   & +0.033 & -0.163 & 165.4508  & -0.063& 164.9392&164.7558\\
   &  87   &  1   & +0.014 & -0.147 & 182.2520  & -0.064& 182.0282&181.8216\\
   & 135   &  1   & +0.038 & -0.052 & 282.6374  & -0.051& 282.3383&282.1326\\
   & 141   &  1   & +0.030 & -0.052 & 295.3159  & -0.052& 295.0944&294.8699\\ \hline
Na &  9    &  1   & +0.315 & -0.202 & 1.9610    & -0.205& 1.9578  &  1.9524\\
   & 15    &  1   & +0.394 & -0.288 & 3.2851    & -0.137& 3.2705  &  3.2664\\
   & 27    &  1   & -0.067 & -0.096 & 5.9634    & -0.114& 5.9532  &  5.9478\\
   & 51    &  1   & -0.078 & -0.081 & 11.3871   & -0.100& 11.3735 & 11.3652\\
   & 59    &  1   & +0.009 & -0.268 & 13.2415   & -0.103& 13.2301 & 13.2197\\
   & 65    &  2   & -0.174 & -0.073 & 14.5532   & -0.105& 14.5438 & 14.5320\\
   & 89    &  1   & -0.016 & -0.083 & 20.0237   & -0.088& 20.0146 & 20.0033\\
   & 113   &  1   & -0.042 & -0.064 & 25.3881   & -0.076& 25.3646 & 25.3538\\ \hline
Cs &  9    &  2   & +0.500 & -0.269 & 1.4594    & -0.206&  1.4577 &  1.4556\\
   & 15	   &  2   & +0.546 & -0.242 & 2.4430    & -0.143&  2.4398 &  2.4381\\
   & 27    &  2   & -0.051 & -0.182 & 4.4394    & -0.121&  4.4341 &  4.4318\\
   & 51    &  1   & -0.116 & -0.067 & 8.4618    & -0.106&  8.4548 &  8.4512\\
   & 59    &  2   & -0.054 & -0.215 & 9.8225    & -0.109&  9.8225 &  9.8181\\
   & 65    &  1   & -0.161 & -0.059 & 10.8073   & -0.105& 10.8011 & 10.7966\\
   & 89    &  2   & -0.035 & -0.066 & 14.8623   & -0.093& 14.8590 & 14.8541\\
   &113    &  2   & -0.013 & -0.143 & 18.8588   & -0.078& 18.8460 & 18.8415\\
\end{tabular}
\end{ruledtabular}
\end{table*}

For comparison, the Table~\ref{tab1} also include the SC-SJM and SJM results. The values $\nu^\dagger=1,2$ imply that the surface region has approximately $t^\dagger=1\times d_{100}$ or $t^\dagger=2\times d_{100}$ thicknesses. The negativity of the values for $\Delta r_s2^\dagger$ implies that SC-ISJM predicts an increase for the ionic density of surface region. However, $\Delta r_s1^\dagger$ values are positive for all Al$_N$, and mostly negative for Na$_N$ and Cs$_N$ clusters. In cases that $\Delta r_s1^\dagger$ are positive, it implies that the inner-region atoms of the corresponding cluster tend to occupy a larger volume, which in turn indicates the expansion of inner region. However, this does not mean the overall expansion of the cluster because the sign of relaxation for the surface region must also be taken into account. These results show the advantages of SC-ISJM over SC-SJM in which all parts of jellium undergo contractions (see column 7 of Table~\ref{tab1}).

In the SC-ISJM, the equilibrium sizes of all clusters, except for Al$_{55}$, are smaller than their un-relaxed values. For Al$_{55}$, the results show an expansion of the jellium. We notice that even though the surface region in this case is contracted ($\Delta r_s2^\dagger$=-0.068), the expansion of the inner region ($\Delta r_s1^\dagger$=+0.206) is dominated and resulted in an overall expansion.

For Al clusters, the SC-ISJM predicts expansions for inner regions, because      
$\Delta r_s1^\dagger >0$ for all of them. In Na$_N$ clusters, the inner regions of all, except for $N=9,15,59$, undergo contractions. For Cs$_N$ clusters, the inner region of only Cs$_9$ and Cs$_{15}$ undergo expansions and the others show contractions.

\begin{figure*}[t]
\begin{center}
\includegraphics[width=0.8\linewidth]{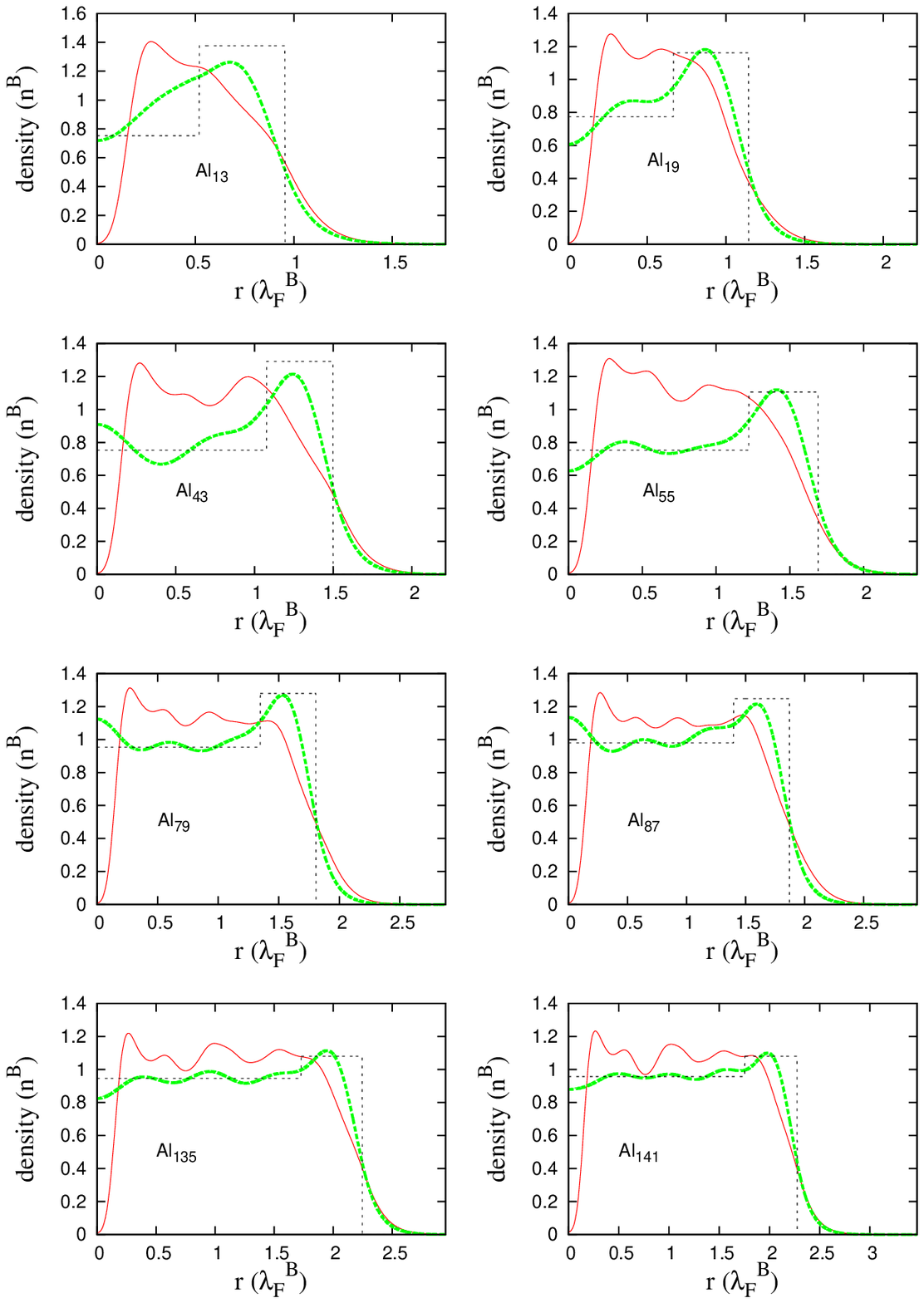}
\caption{\label{fig7} Electronic and ionic jellium densities, in units of bulk density, $n^{\rm B}$, as functions of radial distances for Al$_N$ clusters. Thick dashed green lines and thin dashed black lines correspond to SC-ISJM electronic and jellium densities, respectively. Red solid lines correspond to spherically averaged electron density obtained from first-principles method.} 
\end{center}
\end{figure*}

\begin{figure*}[t]
\begin{center}
\includegraphics[width=0.8\linewidth]{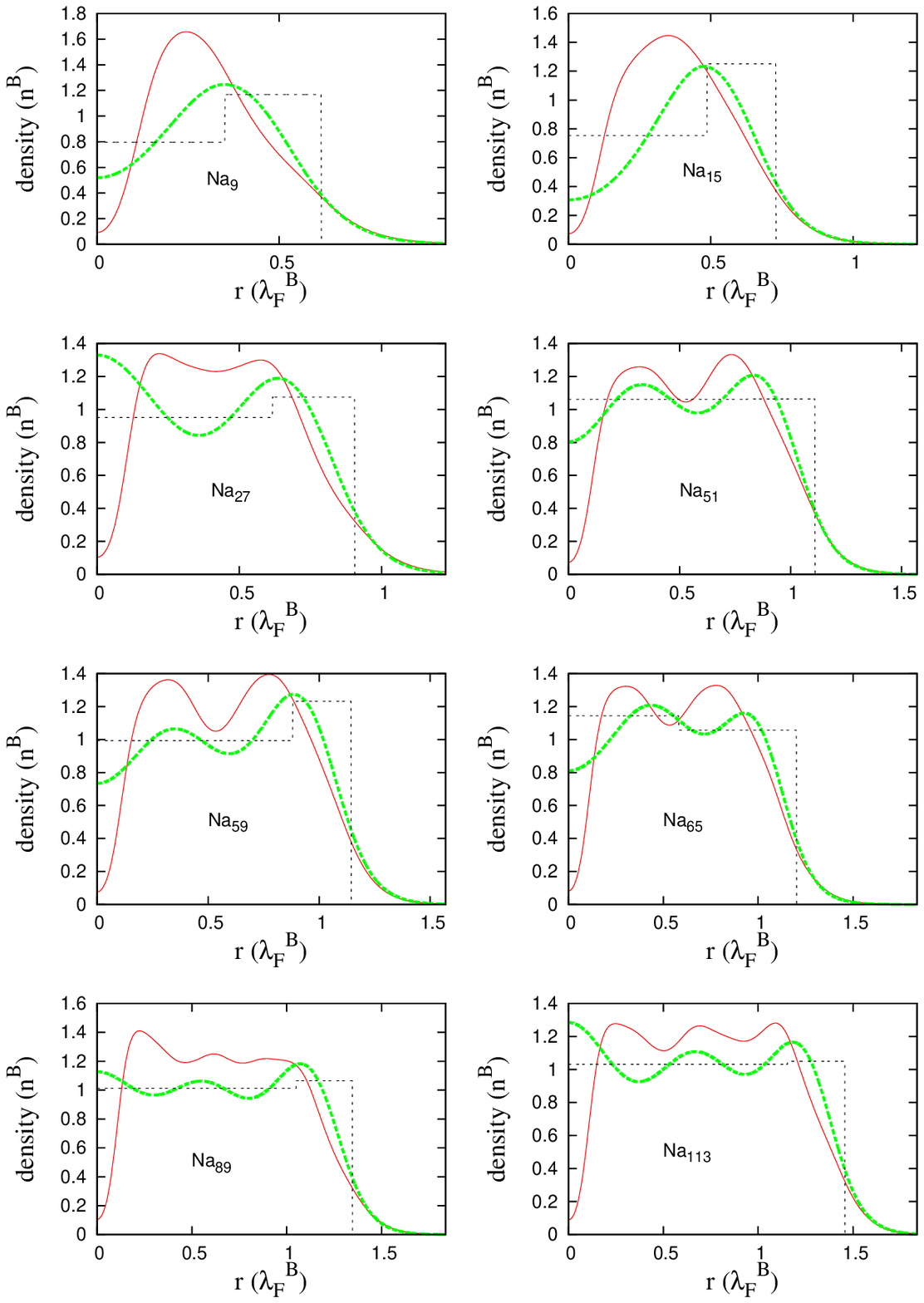}
\caption{\label{fig8} The same as in Fig.~\ref{fig7}, but for Na$_N$ clusters.}
\end{center}
\end{figure*}

\begin{figure*}[t]
\begin{center}
\includegraphics[width=0.8\linewidth]{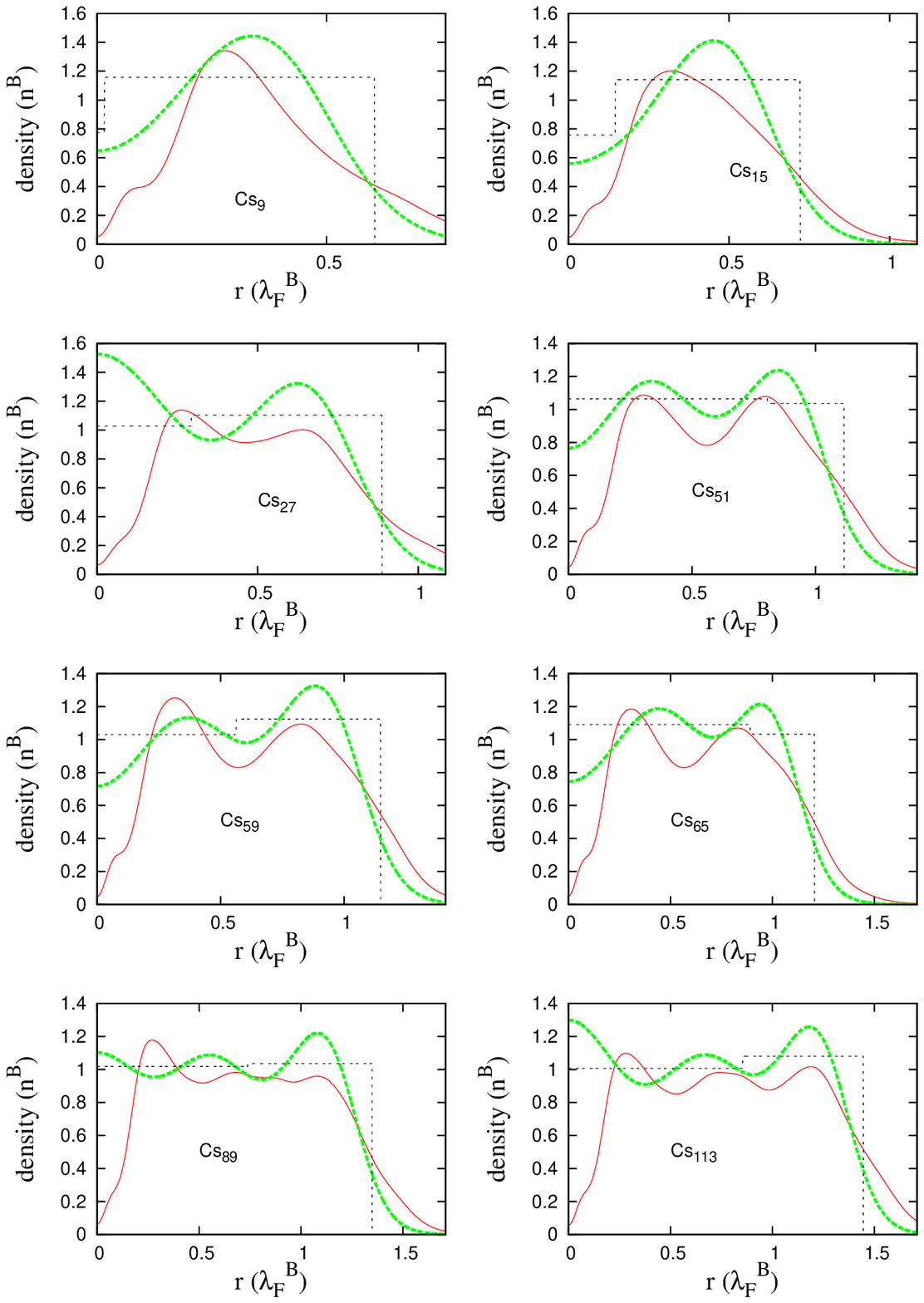}
\caption{\label{fig9} The same as in Fig.~\ref{fig7}, but for Cs$_N$ clusters.}
\end{center}
\end{figure*}

In Figs.~\ref{fig7}-\ref{fig9}, we have plotted the equilibrium SC-ISJM electron and jellium densities of the clusters, and compared with the spherical averaged electron densities of first-principles calculation results. The unit for density is that of bulk jellium, $n^{\rm B}$, and the distances are expressed in terms of Fermi wavelength for bulk jellium, $\lambda_{\rm F}^{\rm B}$. The thick dashed green lines correspond to SC-ISJM electron densities; the thin dashed black lines and solid red lines correspond to jellium and first-principles electron densities, respectively.
Jellium densities of less than unity correspond to expansion with respect to un-relaxed bulk jellium, whereas for larger than unity the system is contracted.

In Fig.~\ref{fig7} which corresponds to Al clusters, we observe that the inner regions are expanded for all clusters, and this expansion is significant for $N$=13, 19, 43, 55. On the other hand, these expansions are accompanied by the contractions of surface regions.

In Na clusters, as Fig.~\ref{fig8} shows, the expansion of inner region is realized for $N$=9, 15, 27 whereas for others it is contracted or remained unchanged. On the other hand, these changes are accompanied by contractions of surface regions. Careful inspection of Fig.~\ref{fig8} corresponding to Na$_{65}$, we notice that even though both surface and inner regions are contracted with respect to the bulk, but the contraction of the surface region is less than that of the inner region. This behavior is in agreement with the relative surface-expansion seen in Fig.~\ref{fig5} corresponding to Na$_{65}$.       

Finally, in Fig.~\ref{fig9}, we have plotted the SC-ISJM results for Cs$_N$ clusters.
For these systems in most cases we have $\nu^\dagger=2$, and for both regions we have jellium contractions, except for $N$=9, 15. Similar to Na$_{65}$, here we have Cs$_{51}$ and Cs$_{65}$ which show relative expansions with respect to inner regions.
Here, the Cs$_{65}$ behavior is in agreement with the first-principles result in Fig.~\ref{fig6} which shows relative expansion of the surface layer.

As to electron densities, the spherical-averaged values of the first-principles calculations are quite similar to the SC-ISJM ones, both in extensions and in the form of Friedel oscillations.\cite{friedel1958metallic} The discrepancy between SC-ISJM and first-principles electron densities near $r=0$ originates from the $r^2$ weight factor which is used in spherical average.

\subsection{Discussion}  
As was mentioned in Section~\ref{sec1}, the JM was developed to describe some electronic properties of metals within a low-cost computational effort. The SJM, which was a mature version of the JM, was able to describe the properties in a broader areas than the JM. However, to describe the geometric properties of finite metals, it was necessary to release the constraint of uniform positive background density, $\bar n^{\rm B}$, and let it take a new value, $\bar n^\dagger$, which stabilizes the system. The latter procedure which gave rise to the SC-SJM, was successful in giving the accurate sizes of simple metal clusters. The next development was to generalize the model in such a way to decouple the relaxations of surface and inner part of the system, which was called the SC-ISJM.

The SC-ISJM calculation results for Al$_N$, Na$_N$, and Cs$_N$ clusters show that all, but Al$_{55}$, undergo size reduction in good agreement with the atomic simulation results. However, in the simulations, the Cs$_{15}$ and Cs$_{65}$ were predicted to increase their sizes. This phenomenon which is shared in both simulation and the SC-ISJM, shows the advantages of the model and some miss-predictions can be attributed to the rough method of choosing the thickness as $t=\nu d_{100}$ or to the fact that assuming ``only two regions'' for the system is possibly not sufficient and one needs to straightforwardly generalize it to three or more regions.

The first-principles results showed that the second shells of most clusters were also contracted. This is also in agreement with the SC-ISJM results because, for some clusters the minimum-energy state occur at $\nu^\dagger$=2 and the sign of $\Delta r_s2^\dagger$ is negative. If in the application of SC-ISJM we would have considered more than two regions, we could then be able to reproduce the relaxations of the third or higher layers as in the simulations. 

Finally, in the first-principles results, the relative distances between the first and second atomic shells were increased for Na$_{15}$, Na$_{65}$, Cs$_{15}$, Cs$_{65}$, and Cs$_{113}$. This relative expansion was also reproduced in the SC-ISJM results for Na$_{65}$, Cs$_{51}$, and Cs$_{65}$, which shows the good agreement of these results with those of atomic simulations.

The good agreement of the SC-ISJM results with those of first-principles method reveals that the model is more realistic than its predecessors and could even be improved. 

\section{\label{sec5}Conclusions}
It is desirable to explain the properties of materials in simple ways by realistic models, because on the one hand it demands much less computational facilities and efforts, and on the other hand it is feasible by a personal computer system. In this respect, the JM and its variants were developed to describe the electronic structure properties of metals. The earlier versions of the JM were utilized to describe the electronic properties of mostly simple metal systems. The newer versions such as SC-SJM and SC-ISJM (subject of present work) can be used to describe some geometric properties either. As was discussed, the SC-SJM was able to predict only the overall size changes, but the SC-ISJM has the ability to reproduce the ionic inhomogeneity of a system.
The SC-ISJM showed that it correctly predicts the size reduction of most clusters of Al$_N$, Na$_N$, and Cs$_N$, as was predicted by {\it ab initio} calculations. On the other hand, we showed that SC-ISJM can also predict the size enlargements in some clusters, as was seen from first-principles results. Finally, the expansions of inter-layer distances which was predicted by the atomic simulations, were correctly reproduced by the SC-ISJM. To overcome some discrepancies observed in the results of the model with those of first-principles method, one should improve the method of choosing surface thicknesses or straightforwardly generalize the model to include more than two regions.  

\begin{acknowledgments}
This work is part of research program in Theoretical and Computational Physics Group, NSTRI, AEOI.
\end{acknowledgments}

\bibliography{Payami-arXiv-09_02_17} 

\end{document}